\documentclass[journal=jpclcd,manuscript=article]{achemso}
\setkeys{acs}{biblabel=brackets}

\usepackage[version=3]{mhchem}

\author{Daniel J. Cole}
\affiliation[University of Cambridge]
{Theory of Condensed Matter Group, Cavendish Laboratory, University of Cambridge, J. J. Thomson Avenue, Cambridge CB3 0HE, UK.}
\author{David D. O'Regan}
\affiliation[\'{E}cole Polytechnique F\'{e}d\'{e}rale de Lausanne]
{Theory and Simulation of Materials, \'{E}cole Polytechnique F\'{e}d\'{e}rale de Lausanne, MXC 341, Station 12, CH-1015 Lausanne, Switzerland.}
\author{Mike C. Payne}
\email{mcp1@cam.ac.uk}
\affiliation[University of Cambridge]
{Theory of Condensed Matter Group, Cavendish Laboratory, University of Cambridge, J. J. Thomson Avenue, Cambridge CB3 0HE, UK.}
\alsoaffiliation[This document is the Accepted Manuscript version of a Published Work that appeared in final form 
\\ \indent \;\; in J. Phys. Chem. Lett., 2012, 3 (11), pp 1448--1452, copyright \copyright \; American Chemical Society 
\\ \indent \;\; after peer review and technical editing by the publisher. 
\\ \indent \;\; To access the final edited and published work see http://pubs.acs.org/doi/full/10.1021/jz3004188.]{}

\title[]
{Ligand Discrimination in Myoglobin from Linear-Scaling DFT+$U$}

\begin{document}

\clearpage

\begin{abstract}

  Myoglobin modulates the binding of diatomic molecules to its heme
  group via hydrogen-bonding and steric interactions with neighboring
  residues, and is an important benchmark for computational studies of
  biomolecules.
  We have performed calculations on the heme binding site and a
  significant proportion of the protein environment (more than 1000
  atoms) using linear-scaling density functional theory and the
  DFT+$U$ method to correct for self-interaction errors associated
  with localized $3d$ states.
  We confirm both the hydrogen-bonding nature of the discrimination
  effect (3.6~kcal/mol) and assumptions that the relative strain
  energy stored in the protein is low (less than 1~kcal/mol).
  Our calculations significantly widen the scope for tackling problems
  in drug design and enzymology, especially in cases where electron
  localization, allostery or long-ranged polarization influence ligand
  binding and reaction.

\end{abstract}

\begin{figure}
\begin{center}
\includegraphics{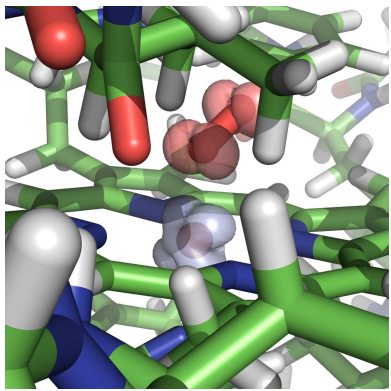}
\caption*{Table of contents graphic.}
\label{fig_si1}
\end{center}
\end{figure}


\clearpage

\begin{figure}
\begin{center}
\includegraphics[width=6.0in]{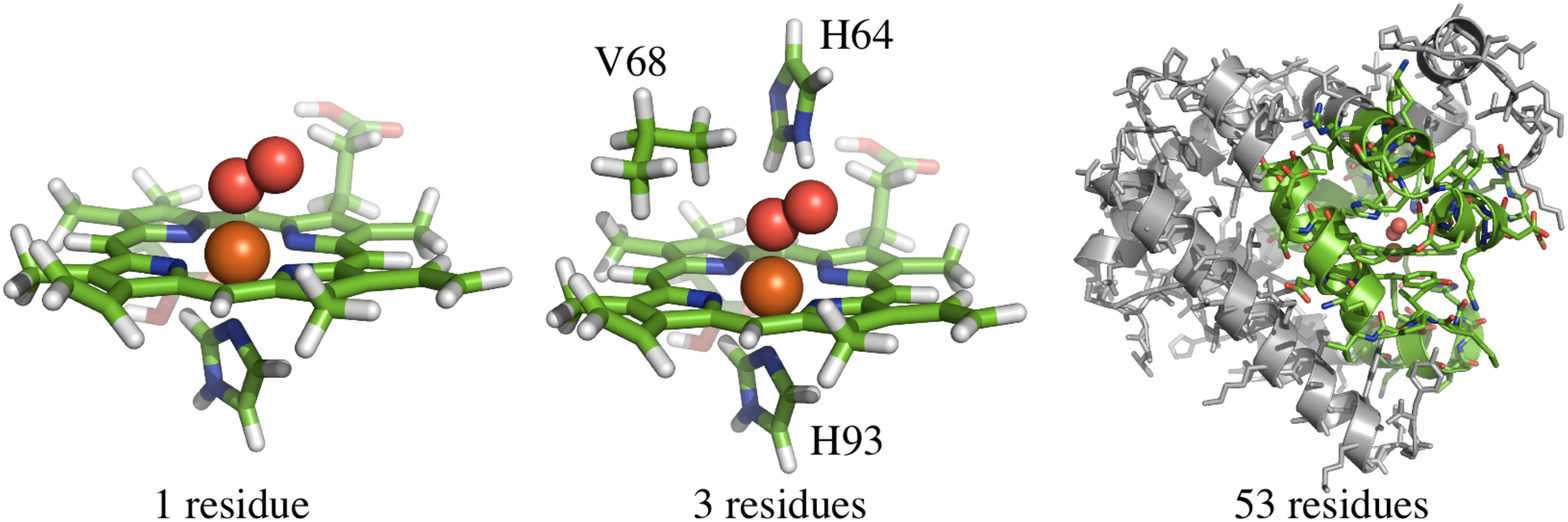}
\caption{Computational models used in the current study. Iron and
  oxygen ligand atoms are represented by orange and red spheres,
  respectively. Distal residues H64 and V68, identified as important in
  determining ligand binding, are labeled, as well as the proximal
  histidine H93. In the 53-residue system, the full myoglobin molecule 
  is shown in grey, while the system used in our calculations is colored.}
\label{fig1}
\end{center}
\end{figure}

Myoglobin (Mb) is a small, globular protein, which is responsible for
storing oxygen in muscle tissues.
Mb contains a single heme group, which is packed within a
predominantly $\alpha$-helical secondary structure and is co-ordinated
by a histidine residue (known as the proximal histidine) as the fifth ligand
of the heme's central iron ion.
The $3d$ electrons of the ferrous heme iron ion (Fe(II)) are
energetically well-aligned with $\pi^*$ acceptor orbitals in CO and
O$_2$ and, as such, are capable of strongly binding these gaseous
molecules.
The Mb protein famously reduces the heme group's natural preference for CO
binding: the binding energy of CO, relative to O$_2$, is reduced
approximately 1000-fold (or $\sim 4$~kcal/mol) in the protein
environment~\cite{olson97}.
The influence of the protein is traditionally split into two
effects~\cite{spiro01}, mediated by two distal protein residues H64
and V68 (\ref{fig1}).
First, the $\pi^*$ acceptor orbitals on O$_2$ are lower in energy than
on CO, resulting in greater charge transfer from the Fe $3d$ orbitals
and, hence, a stronger electrostatic interaction with the neighboring
H64.
Second, the symmetry of unoccupied CO $\pi^*$ acceptor orbitals results
in a linear lowest energy binding conformation with heme: the Fe--C--O
bond angle is close to $180^{\circ}$, while Fe--O--O is closer to
$120^{\circ}$.
It has been argued that steric interactions involving the H64 and V68 residues of
myoglobin reduce the affinity of CO relative to O$_2$, which can be
more easily incorporated into the binding cavity in its lowest energy
bent conformation~\cite{collman76}.

The nature of the protein effect has been well-studied and makes
myoglobin an important benchmark for biocatalysis studies and
structure-function relationships.
Density functional theory (DFT)~\cite{angelis05},
QM/MM~\cite{sigfridsson02} and Poisson-Boltzmann~\cite{phillips99}
studies indicate that the first, hydrogen-bonding factor is the
dominant one in determining preference for O$_2$ binding in myoglobin.
Site-directed mutagenesis of H64 to various hydrophobic residues
implies that it contributes around 3.7~kcal/mol to ligand
discrimination~\cite{springer94}.
In conjunction with DFT studies showing that the strain energy stored
in the CO ligand in more recent crystal structures is quite low
($<1$~kcal/mol)~\cite{spiro01}, a consistent picture of ligand
discrimination via hydrogen-bonding to the distal histidine emerges.
However, as with many similar computational simulations of ligand
binding and reactions in proteins, the protein environment of the heme
group beyond a small number of residues is generally neglected or is
included in a QM/MM description.
This leaves two important questions unanswered.
First, how does long-range polarization by the protein environment
affect charge transfer to the O$_2$ ligand and its electrostatic
interactions with H64?
Second, how much strain energy is stored in the protein itself, and
does this factor affect its ability to discriminate sterically between
ligands?

Here, we address these questions by means of linear-scaling DFT as
implemented in the \textsc{onetep}~\cite{skylaris05} code, which
combines basis set accuracy equivalent to that of plane-wave DFT
methods, with a computational cost that scales linearly with the
number of atoms in the system.
In combination with a DFT energy functional augmented by damped London
terms capable of describing van der Waals interaction in weakly bound
systems (DFT+D)~\cite{hill09}, this method allows for an accurate,
fully QM description of systems of thousands of atoms~\cite{hine09},
including entire proteins~\cite{cole10,cole11}.
However, conventional local or semi-local approximate
exchange-correlation functionals for DFT often fail to describe the
physics of strongly localized orbitals, particularly those of $3d$
symmetry, due to self-interaction errors associated with the lack, or
underestimation, of the exact functional's derivative discontinuity
with respect to particle number~\cite{cohen08}.
Typical symptoms of self-interaction error may include, for example,
underestimated localization and related local moments, spurious
selection of low-spin states, underestimated single-particle and
optical excitation energies, and qualitatively incorrect metal-ligand
binding.
The latter problem, in particular, inhibits the reliable application of such 
functionals to the study of the transition metal binding sites central to the 
function of a range of proteins.
Since conventional, inexpensive functionals often perform acceptably 
well for describing lighter elements, and hybrid functionals comprising a fraction 
of exact exchange are excessively expensive for large systems, we favor the 
DFT+Hubbard $U$ (DFT+$U$, also known as LDA+$U$) approach for correcting 
the functional locally, thus canceling the self-interaction errors directly 
in the $3d$ subspaces where they are most grave.
The DFT+$U$ method is most well-known in solid-state 
physics~\cite{PhysRevB.71.035105,PhysRevB.58.1201}, but has been demonstrated 
to be particularly efficient and effective for correcting the self-interaction 
error in transition metal chemistry~\cite{kulik06,kulik08,kulik11} and is used with increasing 
frequency for biological systems.
The implementation of the DFT+$U$ method in the linear-scaling framework, with 
full optimization of both the local orbitals describing the Kohn-Sham states and 
the states for correction with DFT+$U$, has been recently 
demonstrated~\cite{oregan10,oregan11,oregan12}
and it is with this implementation that we 
perform simulations of a realistic model of the Mb heme protein
(1007 atoms) in complex with two ligands, CO and O$_2$, at a fully
QM level.

In \ref{tab1}, we compare the structural properties of the three
computational models shown in \ref{fig1} extracted from both Mb-CO 
and Mb-O$_2$ X-ray crystal structures~\cite{vojtechovsky99} and optimized 
using DFT, with a Hubbard parameter of $U=0$~eV.
The largest systems require the optimization of 113 atoms within a 894 
atom protein environment (Computational Methods).
The modeled geometries are, in general, in good agreement with each
other and with the experimental structures, and our Mb-O$_2$
geometries compare favorably with optimized DFT/MM structures
employing a range of different exchange-correlation
functionals~\cite{chen08}.
The Fe--C--O bond angle decreases by $\sim5^{\circ}$ on the addition of
the neighboring protein residues, H64 and V68, while the O$_2$
molecule retains a bent conformation of $\sim120^{\circ}$.
The root mean square deviation (RMSD) of the computed atomic positions of the heme
structures from those in the experimental structures are low and most of the
RMSD increase in the 53-residue system is due to the propionate side
chains.
Residue H64 is observed to adopt multiple conformations in crystal structures
of Mb-O$_2$~\cite{vojtechovsky99,unno07}.
We have found the more distant conformation (O--N$_{{\rm H64}}$ distance of
2.97~\AA{}) to be the more energetically favorable conformation in the
3-residue Mb-O$_2$ simulation (by 1.9~kcal/mol) and results reported in this letter 
are specifically for this conformation.
Following complete relaxation in the 53-residue system, the O--N$_{{\rm H64}}$ distance is
intermediate between the two reported experimental values~\cite{vojtechovsky99}.
\begin{table}
\begin{center}
\begin{tabular}{lccccccccc}
 & \multicolumn{4}{ c }{Mb-CO}  &&  \multicolumn{4}{ c }{Mb-O$_2$} \\
\cline{2-5}\cline{7-10} \\
 & 1 & 3 & 53 & 1A6G && 1 & 3 & 53 & 1A6M\\
\hline
Fe -- X               & 1.79 & 1.78 & 1.77 & 1.82 && 1.92 & 1.87 & 1.86 & 1.81 \\
X -- O                & 1.16 & 1.16 & 1.16 & 1.09 && 1.24 & 1.26 & 1.27 & 1.24 \\
X -- N$_{{\rm H64}}$     & --   & 3.47 & 3.46 & 3.42 && --   & 3.09 & 3.01 & 3.08/3.02 \\
O -- N$_{{\rm H64}}$     & --   & 3.25 & 3.33 & 3.16 && --   & 2.87 & 2.78 & 2.97/2.67 \\
Fe -- N$_{{\rm H93}}$    & 2.08 & 2.09 & 2.10 & 2.06 && 2.14 & 2.08 & 2.06 & 2.06 \\
Fe -- N$_{{\rm heme}}$   & 2.03 & 2.04 & 2.03 & 1.99 && 2.03 & 2.03 & 2.03 & 2.01 \\
Fe -- X -- O          & 175.0$^{\circ}$& 170.2$^{\circ}$& 170.3$^{\circ}$& 171.1$^{\circ}$&& 120.9$^{\circ}$& 120.3$^{\circ}$& 120.4$^{\circ}$ & 122.5$^{\circ}$\\
\hline
RMSD                  & 0.07&0.07&0.11& --    && 0.07&0.06&0.10& -- \\
\hline
\end{tabular}
\end{center}
\caption{Structural data for computational models, with an increasing number of protein
  residues, together with experimental structures. 
  X=C,O for Mb-CO and Mb-O$_2$ respectively. Histidine 
  residues are labeled in \ref{fig1}. The root mean square 
  deviation of the computed atomic positions from those in the experimental structures 
  (RMSD) is measured for non-hydrogen atoms of the 
  heme group and ligand. 
  Two alternative ligand to N$_{{\rm H64}}$ distances are quoted for the 1A6M 
  crystal structure, corresponding to two equally occupied conformations of the distal 
  histidine. Distances are measured in \AA{}.~\label{tab1}} 
\end{table}

The strength of the hydrogen bond between the distal histidine H64 and
the two ligands is expected to depend strongly on the charge transfer
from the heme group to the ligand.
\ref{tab2} reveals that the magnitude of the natural bond orbital
charge population~\cite{reed85} on CO is less than 0.1~$e$ in all
models studied.
In contrast, there is substantial charge transfer to the O$_2$ ligand,
which increases with system size and reaches $-0.46$~$e$ in the largest
system, in agreement with previous CASSCF/MM calculations~\cite{chen08}.
\begin{table}
\begin{center}
\begin{tabular}{lcccccc}
 & \multicolumn{2}{ c }{Mb-CO}  &&  \multicolumn{2}{ c }{Mb-O$_2$} \\
\cline{2-3}\cline{5-6} \\
 & Fe & CO && Fe & O$_2$\\
\hline
1   & 0.95 & -0.05 && 1.21 & -0.30\\
3   & 0.96 & -0.09 && 1.13 & -0.37\\
53  & 0.39 &  0.01 && 0.82 & -0.46\\
\hline
\end{tabular}
\end{center}
\caption{Charges on the Fe ion and ligand calculated using natural 
  population analysis, expressed in 
  units of the electronic charge, calculated at $U=0$~eV. 
  Results vary by less than 0.06 $e$ with the application of the Hubbard $U$ (up to 5~eV)
  and the Mulliken charges show the same dependency on ligand type and system size 
  (Table S1).~\label{tab2}}
\end{table}

\begin{figure}
\begin{center}
\includegraphics[width=6.0in]{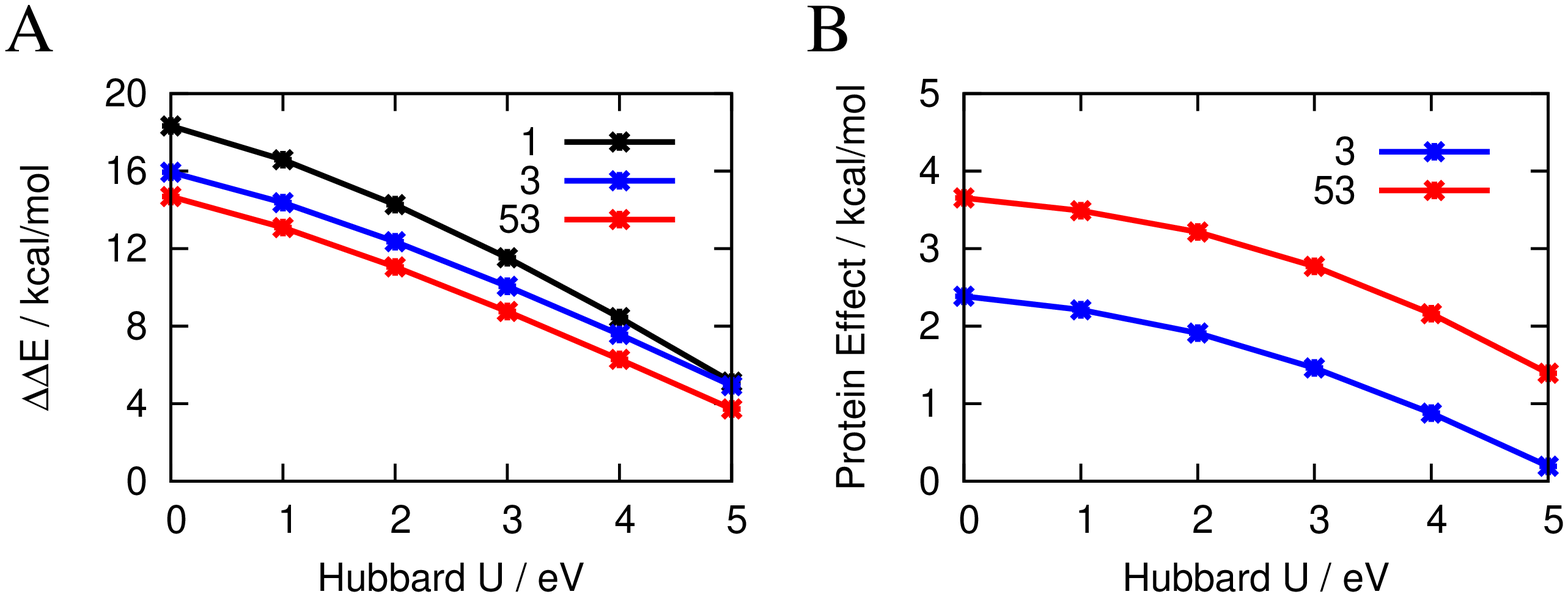}
\caption{(A) Relative binding energies of O$_2$ and CO to heme
  models containing increasing numbers of protein residues (positive
  values indicate that CO binding is favored). The protein decreases
  the preference for CO binding, as expected. (B) The magnitude of
  this protein effect on ligand discrimination as a function of the
  Hubbard $U$ parameter. Dashed lines are estimated from experimental equilibrium
  constants~\cite{olson97}.}
\label{fig2}
\end{center}
\end{figure}

If the dominant ligand discrimination mechanism in myoglobin were due 
to ligand--H64 hydrogen-bonding, then we would expect the larger model 
proteins, in which the charge transfer to O$_2$ is greater, to discriminate 
more strongly in favor of O$_2$.
Indeed, in the absence of the DFT+$U$ correction, the relative affinity
of the heme group for CO is reduced in the 3-residue model, and
reduced further in the 53-residue model (\ref{fig2}A).
The protein effect, defined as the difference between the
relative binding energies of CO and O$_2$ to the heme group calculated 
in vacuum and within the protein model system, is
3.7~kcal/mol and 2.4~kcal/mol in the 53-residue and 3-residue systems
respectively.
Although there is no equivalent experimental probe of the effect of
the protein on the relative enthalpies of binding that we have
measured here, the effect of the protein on the relative free energy
of binding is inferred to be $\sim4$~kcal/mol from equilibrium
constants~\cite{olson97}.
A rigorous comparison between the two results would require estimates
of the effect of the protein on the relative entropy of binding of the
two molecules, which is usually assumed to be small, and solvation
effects, though the heme binding site would seem to be well-isolated
from any electrostatic interaction with the solvent.
A promising avenue for further work is the investigation of the
effects of finite temperature sampling of protein side chain
conformations on the relative binding
energies.~\cite{alcantara07,strickland06}
However, encouraged by the similarity between our results and the
experimental protein effect subject to the caveats discussed, we make
the assumption that our static ground state calculations are
indicative of the finite temperature discrimination
mechanism~\cite{angelis05}.

In order to examine the discrimination mechanism, we have decomposed
the protein effect into distal interaction and strain energies that
involve the protein fragment containing the residues H64 and V68
(Supporting Information).
The first effect is the distal relative interaction energy between the 
heme-ligand complex and the protein fragment containing H64 and V68 
and reflects the overall trend in discrimination -- 
these interactions favor O$_2$ binding by 3.6~kcal/mol in the 
53-residue system and by just 1.7~kcal/mol in the minimal 3-residue system.
These results are consistent with experimental~\cite{springer94} and
DFT-simulated~\cite{angelis05} site-directed mutagenesis of H64, which
imply that the hydrogen bond contributes 3.7 or 3.3~kcal/mol,
respectively, to ligand discrimination, and with QM/MM simulations~\cite{sigfridsson02},
which give the difference in hydrogen bond strengths to be 5~kcal/mol.
The second effect -- the distal relative strain energy stored in the
protein fragment (the energy difference between the optimized Mb-CO
and Mb-O$_2$ structures with the heme group, ligand and the remainder
of the protein removed) -- is more difficult to estimate using
conventional, cubic-scaling DFT or QM/MM calculations, requiring a
method sensitive to atomic displacements over a large region.
In the 3-residue model, H64 and V68 store 0.3~kcal/mol of strain energy 
when binding CO (relative to O$_2$).
By fully optimizing the heme group and surrounding residues within the
constraints of the full protein environment with first principles QM,
we find that the distal relative strain falls to $-0.1$~kcal/mol in
the 53-residue system, implying that the strain is not only dissipated
by allowing rearrangement of the surrounding matrix, but also that
there is a slight steric preference for CO binding.
This implies that formation of the H64--O$_2$ hydrogen bond may incur 
a strain penalty in the protein matrix, an effect which is not measurable
using the minimal model.
In both cases, however, the energy stored is much smaller than the
polar discrimination effect and, as is generally assumed~\cite{spiro01}, 
is of the same order as the CO distortion energy.
It should be emphasized that, within the 53-residue model, only the 
heme group, ligand and three residues around the binding site have been 
re-optimized upon CO binding, thus concentrating the strain energy 
difference to this local region.
However, extending the relaxations to allow the optimization of eight
residues (H64--V68 and S92--A94) gave no change in the distal relative
interaction energy (3.6~kcal/mol) and a distal relative strain of
$-0.3$~kcal/mol, implying that both quantities are converged.
The likelihood is that this optimization region would need to be extended 
in less rigid protein structures, though the details of the convergence of the 
elastic energy with system size is beyond the scope of this work.

Despite the agreement between the nature of the protein effect
calculated here and the literature consensus, little is known about
how the protein responds to the treatment of electron localization at
the metal center.
For example, the tendency of local exchange-correlation functionals to
over-delocalize electrons in correlated systems leads to
over-estimation of metal-ligand binding energies.
In agreement with previous DFT+$U$ calculations~\cite{scherlis07}, we
find that the addition of the localized self-interaction correction
recovers the quintuplet electronic ground state of the isolated heme
group and improves the local magnetic moments and, to some extent,
Fe-ligand binding energies (Figures S1 and S2).
With this improved description of the Fe $3d$ orbital occupancies, the
relative affinity of heme for CO is reduced, both with and without the
protein environment (\ref{fig2}A), and the protein effect,
correspondingly, decreases somewhat with increasing $U$ (\ref{fig2}B).
However, this variation is all in the treatment of the heme/ligand sub-system, 
as expected from an intrinsically short-ranged correction.
The distal relative interaction energy varies much less (between 3.6 and 
4.3~kcal/mol, Figure S3) and the distal relative strain is, by definition, 
unchanged.

We have demonstrated that linear-scaling DFT may be applied to a heme
group embedded in a realistic model of the heme protein, consisting of
over 1000 atoms, and used to elucidate features of a much-discussed
discrimination mechanism.
Compared to the minimal model of the heme-protein interaction, the
53-residue model of Mb-O$_2$ has a larger charge transfer to the
ligand, which results in a stronger hydrogen-bonding interaction with
the distal side of the protein.
In comparison, the relative strain energy stored in the protein is
smaller, and is certainly not enough to account for the ligand
discrimination effect in myoglobin.
While a systematic study of the convergence of energetic properties of
heme with system size, the finite temperature sampling of protein side
chain conformations, and a complete description of correlated electron
effects in ligand-heme interactions, remain as future avenues of
investigation, our work represents an encouraging, positive result for
the feasibility of the use of large-scale DFT simulations in a
well-studied benchmark system.
The treatment of entire proteins at the full QM level is now becoming
widespread~\cite{cole10,cole11,ufimtsev11}, potentially increasing the
accuracy and range of problems that are open to study in fields ranging 
from small molecule therapeutics to enzymology.

\section{Computational Method}

QM calculations were performed with spin-polarized DFT as implemented
in the \textsc{onetep} code~\cite{skylaris05}, using the PBE gradient
corrected exchange-correlation functional~\cite{perdew96}.
Van der Waals interactions were approximated by augmenting the DFT
energy functional with damped London potentials with parameters
optimized specifically for the PBE functional~\cite{hill09}.
The \textsc{onetep} parameters used are described in the Supporting
Methods.
Computational models of the heme site were based on X-ray crystal
structures of sperm whale Mb in two different ligation states:
oxygenated (Mb-O$_2$, PDB: 1A6M) and
carbonmonoxygenated (Mb-CO, PDB: 1A6G)~\cite{vojtechovsky99}.
Three different sized systems were constructed by retaining one (H93),
three (H64, V68 and H93) or 53 (I28--F46, K63--L72, P88--I111)
residues surrounding the heme group in the two X-ray crystal structures.
Geometry optimization~\cite{hine11} was performed to tolerances of
0.003~eV and 0.05~eV/\AA{} on the total energies and forces
respectively.
Initial geometry optimization of the 53-residue Mb-O$_2$ structure
revealed substantial spin populations on four charged residues on the
exterior of the model system (K34, E41, D44 and K63).
These residues would either form salt-bridges or would be
solvent-exposed in their real environment, but are exposed to the
vacuum in our model system.
We therefore manually transferred protons from K34 and K63 to E41 and D44
in order to neutralize these residues and reduce the 
spurious local magnetic moments to zero.
Following further optimization to a force tolerance of 0.2~eV/\AA{},
the outer 50 residues were frozen and the heme group and three closest
residues (H64, V68 and H93) were optimized to the tolerances described
above.
Mb-CO was modeled by replacing the heme group and residues H64, V68
and H93 by their positions in the carbonmonoxygenated crystal
structure, and allowing them to relax in the identical frozen 50-residue
environment.
This scheme ensures that energy differences are directly attributable to 
local changes in the binding site, whilst accounting for long-ranged 
polarization and constraints of the protein scaffold.
It is important, however, to ensure that results are converged with 
respect to the size of the relaxed region.

To correct for the underestimation of $3d$ electron localization 
by the approximate exchange-correlation functional, a 
rotationally-invariant form of the widely-used DFT+$U$ 
correction~\cite{PhysRevB.71.035105,PhysRevB.58.1201} was
applied to the Fe $3d$ manifold. 
In this  method, implemented in the linear-scaling DFT 
formalism~\cite{oregan10,oregan11,oregan12}, and 
previously applied to heme systems with 
success~\cite{PhysRevB.79.245404,scherlis07}, 
an additional term is added to the
energy functional, taking the form for a single Fe ion:
\begin{equation}
E_U = \frac{U}{2} \sum_\sigma \sum_m 
\left[ n^\sigma_{mm} - \sum_{m'} n^\sigma_{m m'} n^\sigma_{m' m} \right],
\quad \mbox{where} \quad
n^\sigma_{m m'} = \sum_i \langle \varphi_m \rvert \psi^\sigma_i \rangle f^\sigma_i
\langle \psi^\sigma_i \lvert \varphi_{m'} \rangle.
\end{equation}
Here, the functions $\lvert \psi^\sigma_i \rangle $ are Kohn-Sham orbitals of
spin $\sigma$ and occupancy $f^\sigma_i$.
The $\lvert \varphi_m \rangle$ are the Hubbard projectors that
delineate the subspace for correction, in this case, iron-centered
atomic $3d$ orbitals that are numerically solved using the
corresponding atomic pseudopotential.
The Hubbard $U$ determines the strength of the DFT+$U$ correction,
and while it may be computed from first principles
using a number of different methods, it is treated as a 
free parameter in our calculations.
The net effect, with increasing $U$, is to more strongly 
penalize deviation from integer values of the eigenvalues
of the $3d$ subspace occupancy matrices $n^\sigma_{m m'} $
and correct for localized self-interaction errors in the 
approximate functional~\cite{PhysRevB.71.035105,PhysRevB.58.1201}.
The DFT+$U$ ground state wave-function remains single-determinantal,
with no finite temperature or explicitly dynamical quantum effects
included, although the extension of DFT+$U$ to dynamical mean-field
theory, offering improved descriptions of excited-state quantities
such as optical spectra, has recently been applied to heme
(unpublished results).
In order to describe the dominant open-shell singlet ground state 
of Mb-O$_2$ within Kohn-Sham DFT, it was necessary to break the
magnetic symmetry by applying effective magnetic fields, of opposite 
sign, locally to the Fe $3d$ and O$_2$ $2p$ manifolds.
These effective fields were implemented by means of an extension of
the DFT+$U$ functionality, and were switched
off automatically when the electronic density began to converge. 
The independence of the total energy with respect to 
the strength of the applied fields was carefully tested.
All calculations at DFT+$U>0$~eV were performed on converged
DFT+$U=0$~eV structures, which introduced an estimated error 
of less than 0.2~kcal/mol (Table S2).
%

%% classes.  Rather than use \section, an appropriate macro is
%% provided that will always work.
%%%%%%%%%%%%%%%%%%%%%%%%%%%%%%%%%%%%%%%%%%%%%%%%%%%%%%%%%%%%%%%%%%%%%
\acknowledgement

We are grateful to Nicholas Hine for helpful discussions.
Computational resources were provided by the Cambridge HPC Service,
funded by EPSRC Grant EP/F032773/1.
DJC, DDO'R and MCP acknowledge support from the EPSRC.
%

%% \suppinfo macro.
%%%%%%%%%%%%%%%%%%%%%%%%%%%%%%%%%%%%%%%%%%%%%%%%%%%%%%%%%%%%%%%%%%%%%
%%\suppinfo
\begin{suppinfo}
  Supporting methods, calculation of distal interaction and strain
  energies, Mulliken charge analysis, energetics of FeP(Im), variation
  of local magnetic moments and distal interaction energy with $U$,
  and tests of our geometry optimization protocol at non-zero Hubbard
  $U$.
\end{suppinfo}

\providecommand*\mcitethebibliography{\thebibliography}
\csname @ifundefined\endcsname{endmcitethebibliography}
  {\let\endmcitethebibliography\endthebibliography}{}

\end{document}

\clearpage

\section{Supporting Methods}

\subsection{Computational Models}

Computational models of the heme site were extracted from X-ray crystal
structures of sperm whale Mb in oxygenated (Mb-O$_2$, PDB: 1A6M) and
carbonmonoxygenated (Mb-CO, PDB: 1A6G) ligation states.
Waters of crystallization and sulfate ions were removed.
Oxygenated and carbonmonoxygenated 1- and 3-residue systems were
extracted from the relevant crystal structure and modelled with the
propionate side chains hydrogen-terminated, giving zero net charge.
Histidine and valine residues were terminated at the C$_\gamma$ and
C$_\beta$ atoms, respectively, and these atoms were held fixed at
their crystal positions in the 3-residue simulation in order to model the
constraints of the protein environment.
The 53-residue system was extracted from the Mb-O$_2$ crystal structure,
with protein chains terminated by acetyl and methylamino groups (using
the \textsc{amber10} package~\cite{case08}) and with all amino acids
except histidine in their normal protonation states at pH~7.
The assignment of histidine protonation states were taken from the
literature~\cite{strickland06} and, in particular, H64 was modelled as
the $\epsilon$-tautomer.
The resulting complex consisted of a total of 1007 atoms, with a net
charge of $+2 e$.

\subsection{\textsc{onetep} Parameters}

The \textsc{onetep} program achieves computational cost that scales
linearly with the number of atoms by exploiting the
``near-sightedness'' of the single-particle density matrix in
non-metallic systems~\cite{prodan05}, though no density matrix
truncation was applied in our calculations in order to maximize accuracy.
The density matrix is expressed in terms of a set of nonorthogonal
generalized Wannier functions (NGWFs)~\cite{skylaris02} that are
localized in real space, with truncation radii set to 4.8~\AA{}, and 
optimized during the process of total energy minimization.
The NGWFs were initialized to numerical orbitals 
obtained from solving the Kohn-Sham
equation for free atoms, with a $1s$ configuration for H, a $2s2p$
configuration for C, N and O, and a $3d4s4p$ configuration for
Fe~\cite{serrano-pulay}.
The NGWFs were expanded in a basis of periodic cardinal sine (psinc)
functions~\cite{mostofi03} with a kinetic energy cut-off of 920~eV.
The protein effect was estimated to be converged to less than 0.5~kcal/mol 
with respect to changes in NGWF radii and psinc energy cut-off.
Core electrons were described using norm-conserving pseudopotentials,
including a non-linear core correction for that of the Fe ion.
The spherical cut-off approach for Coulomb potentials was used to
eliminate all interactions of the molecules with their periodic
images~\cite{hine11-2}.
Dispersion interactions were included by augmenting the DFT energy
expression by damped London potentials with parameters optimized
specifically for the PBE functional~\cite{hill09}.
The root mean square error in gas phase binding energies of a
benchmark set of complexes calculated using the DFT methodology
described above has been shown to be approximately 1~kcal/mol when
compared to MP2 and CCSD(T) methods extrapolated to the complete basis
set limit~\cite{hill09}.
Atomic charges and spin populations were calculated using Mulliken and
natural bond orbital analyses implemented within
\textsc{onetep}~\cite{reed85,lee12}.

\section{Distal Interaction Energy and Strain}

The protein effect on the relative binding energies of CO and O$_2$ 
to the heme group consists of: direct interactions between the heme-ligand 
subsystem and the protein, distortions of the protein by the heme-ligand 
subsystem and distortions of the heme-ligand subsystem by the protein.
The protein effect in Figure 2B contains all of these effects.
The last contribution, describing changes in transition metal-ligand geometries, 
is the least accurate (even with the Hubbard $U$ correction), but also the 
most studied using minimal models and more accurate descriptions of correlated
electron effects~\cite{chen08,forthcominghemedmft}.
For these reasons, we have decomposed the protein effect to describe the 
first two contributions.
Namely, the distal relative interaction energy is:
\begin{equation}
\Delta\Delta E = \Delta E({\rm total}) - \Delta E({\rm K63-L72}) - \Delta E({\rm heme, ligand, I28-F46, P88-I111}).
\end{equation}
$\Delta E({\rm total})$ is the difference in total energy between the optimized CO and O$_2$ 
complexes.
K63-L72 is the protein fragment containing distal residues H64 and V68 and the 
remainder of the complex is described by the last term in the equation.
We choose the distal interaction energy so that it may be compared with 
experimental H64 mutagenesis studies and to remove inaccuracies in the 
description of the Fe--H93 interaction.
Finally, the distal relative strain is defined as $\Delta E({\rm K63-L72})$.
\\
\\
\\

\begin{table}
\begin{center}
\begin{tabular}{lcccccc}
 & \multicolumn{2}{ c }{Mb-CO}  &&  \multicolumn{2}{ c }{Mb-O$_2$} \\
\cline{2-3}\cline{5-6} \\
 & Fe & CO && Fe & O$_2$\\
\hline
1   & 1.20 & -0.12 && 1.27 & -0.26\\
3   & 1.17 & -0.16 && 1.19 & -0.41\\
53  & 0.22 & 0.06  && 0.79 & -0.58\\
\hline
\end{tabular}
\end{center}
\caption*{Table S1. Charges on the Fe ion and ligand from Mulliken analysis 
  in the NGWF basis in units of the electronic charge.~\label{tab_si3}}
\end{table}
\begin{figure}
\begin{center}
\includegraphics[width=3.0in]{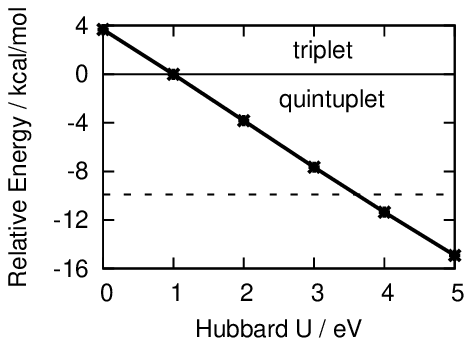}
\includegraphics[width=3.0in]{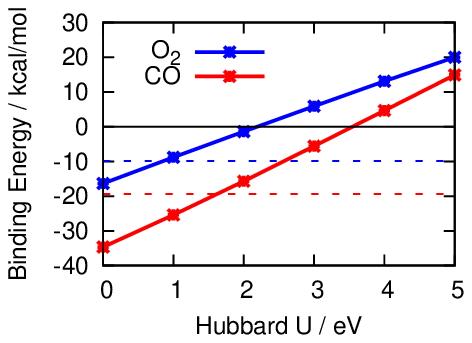}
\caption*{Figure S1. (left) Relative energies of triplet and
  quintuplet spin states of FeP(Im) extracted from the unligated X-ray
  crystal structure (PDB: 1A6N~\cite{vojtechovsky99}). The
  experimental quintuplet ground state is favored at $U>1$~eV. (right)
  Binding energies of O$_2$ and CO ligands to FeP(Im) in
  vacuum. Binding is over-estimated in the absence of the DFT+$U$
  correction. Dashed lines show the respective CASPT2
  results~\cite{radon08}.}
\label{fig_si2}
\end{center}
\end{figure}
\begin{figure}
\begin{center}
\includegraphics[width=3.0in]{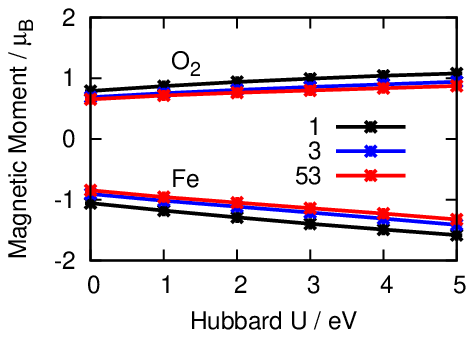}
\includegraphics[width=3.0in]{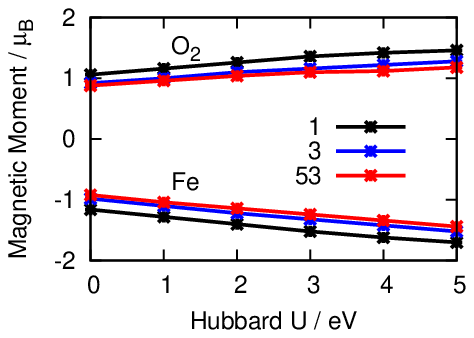}
\caption*{Figure S2. Local magnetic moments on the iron and O$_2$ species in models
  of increasing numbers of residues, consistent with an open-shell
  singlet spin state. The moments are calculated using 
  (left) the DFT+$U$ occupancy matrices (Equation 1) and (right) Mulliken analysis in the NGWF basis.}
\label{fig_si3}
\end{center}
\end{figure}
\begin{figure}
\begin{center}
\includegraphics[width=3.0in]{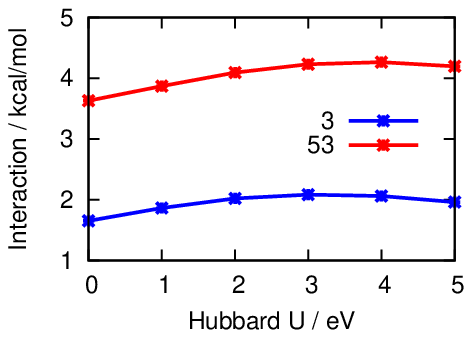}
\caption*{Figure S3. Variation of distal relative interaction energy
  with the DFT+$U$ parameter.  The former is mostly long-ranged, while
  the latter determines the strength of an intrinsically short-ranged
  self-interaction correction.}
\label{fig_si4}
\end{center}
\end{figure}
\begin{table}
  \caption*{Table S2. Relative binding energies of CO and O$_2$ to two heme models with increasing numbers of protein residues 
    ($\Delta\Delta$E), at $U=3.0$~eV, for two different protocols. \emph{Geometry optimize} indicates that the structures 
    have been re-optimized to a force tolerance of 0.15~eV/\AA{}, at $U=3.0$~eV, and 
    \emph{single point} indicates that a single calculation with $U=3.0$~eV has been performed on the structure optimized at $U=0$~eV. 
    The maximum error introduced by not re-optimizing is less than 0.2~kcal/mol.~\label{tab_si2}}
\begin{center}
\begin{tabular}{lcccc}
 &  \multicolumn{2}{ c }{$\Delta\Delta$ E / kcal/mol} \\
\cline{2-3} \\
& 1 residue & 3 residues & Protein Effect\\
\hline
Geometry Optimize & 11.47 & 9.88 & 1.59\\
Single Point      & 11.52 & 10.06 & 1.46\\
\hline
\end{tabular}
\end{center}
\end{table}

\clearpage

\providecommand*{\mcitethebibliography}{\thebibliography}
\csname @ifundefined\endcsname{endmcitethebibliography}
{\let\endmcitethebibliography\endthebibliography}{}